\documentstyle[aps,preprint,epsf]{revtex}	
\def \ppb      {\relax\ifmmode{p\bar p}\else{$p\bar p$}\fi}
\def \chichi   {\relax\ifmmode{\tilde{\chi}_1^\pm \tilde{\chi}_2^0}
                \else{${\tilde{\chi}_1^\pm \tilde{\chi}_2^0}$}\fi}
\def \chizero  {\relax\ifmmode{\tilde{\chi}_1^0}\else{$\tilde{\chi}_1^0$}\fi}
\def \chione   {\relax\ifmmode{\tilde{\chi}_1^\pm}
                \else{$\tilde{\chi}_1^\pm$}\fi}
\def \chitwo   {\relax\ifmmode{\tilde{\chi}_2^0}\else{$\tilde{\chi}_2^0$}\fi}
\def \dk       {\relax\ifmmode{\rightarrow}\else{$\rightarrow$}\fi}
\def \lnu      {\relax\ifmmode{\ell^{\pm}\nu}\else{$\ell^{\pm}\nu$}\fi}
\def \psp      {\relax\ifmmode{\,}\else{$\,$}\fi}
\def \lplm     {\relax\ifmmode{\ell^{+}\ell^{-}}\else{$\ell^{+}\ell^{-}$}\fi}
\def \sp       {\relax\ifmmode{\;}\else{$\;$}\fi}
\def \mgev     {GeV/$c^{2}$}
\def \invpb    {\relax\ifmmode{\rm pb^{-1}}\else{$\rm pb^{-1}$}\fi}
\def \PT       {\relax\ifmmode{P_{T}}
                \else{$P_{T}$}\fi}
\def \pgev     {GeV/$c$}
\def \ET       {\relax\ifmmode{E_{T}}
                \else{$E_{T}$}\fi}
\def \epem     {\relax\ifmmode{e^+e^-}\else{$e^+e^-$}\fi}
\def \upum     {\relax\ifmmode{\mu^+\mu^-}\else{$\mu^+\mu^-$}\fi}
\def \Zz       {\relax\ifmmode{Z^0}\else{$Z^0$}\fi}
\def \Wpm      {\relax\ifmmode{W^{\pm}}\else{$W^{\pm}$}\fi}
\def \ttb      {\relax\ifmmode{t\bar t}\else{$t\bar t$}\fi}
\def \Wp       {\relax\ifmmode{W^{+}}\else{$W^{+}$}\fi}
\def \Wm       {\relax\ifmmode{W^{-}}\else{$W^{-}$}\fi}
\def \gluino   {\relax\ifmmode{\tilde{g}}\else{$\tilde{g}$}\fi}
\def \squark   {\relax\ifmmode{\tilde{q}}\else{$\tilde{q}$}\fi}
\def \stop     {\relax\ifmmode{\tilde{t}}\else{$\tilde{t}$}\fi}
\def \gtsim    {\relax\ifmmode{\mathrel{\mathpalette\oversim >}}
                  \else{$\mathrel{\mathpalette\oversim >}$}\fi}
\def \ltsim    {\relax\ifmmode{\mathrel{\mathpalette\oversim <}}
                  \else{$\mathrel{\mathpalette\oversim <}$}\fi}
\def\oversim#1#2{\lower4pt\vbox{\baselineskip0pt \lineskip1.5pt
            \ialign{$\mathsurround=0pt#1\hfil##\hfil$\crcr#2\crcr\sim\crcr}}}
\def \intlum   {\int {\cal L} dt}
\def \ee       {\relax\ifmmode{e^+e^-}\else{$e^+e^-$}\fi}
\def \mumu     {\relax\ifmmode{\mu^+\mu^-}\else{$\mu^+\mu^-$}\fi}
\font\eightit=cmti8
\def\r#1{\ignorespaces $^{#1}$}
\begin{document}
\draft
\title{
\begin{flushright}
CDF/PUB/EXOTIC/CDFR/3440 \\
FERMILAB-PUB-96/029-E \\[.2in]
\end{flushright}
Search for Chargino-Neutralino Production
in \ppb\sp Collisions at $\sqrt{s} = 1.8$ TeV}
\vspace{.2in}
\author{
\hfilneg
\begin{sloppypar}
\noindent
F.~Abe,\r {14} H.~Akimoto,\r {32}
A.~Akopian,\r {27} M.~G.~Albrow,\r 7 S.~R.~Amendolia,\r {23} 
D.~Amidei,\r {17} J.~Antos,\r {29} C.~Anway-Wiese,\r 4 S.~Aota,\r {32}
G.~Apollinari,\r {27}
% ...
\end{sloppypar}
\vskip .025in
\begin{center}
(CDF Collaboration)
\end{center}
\vskip .025in
\begin{center}
\r 1  {\eightit Argonne National Laboratory, Argonne, Illinois 60439} \\
\r 2  {\eightit Istituto Nazionale di Fisica Nucleare, University of Bologna,
I-40126 Bologna, Italy} \\
\r 3  {\eightit Brandeis University, Waltham, Massachusetts 02254} \\
\r 4  {\eightit University of California at Los Angeles, Los 
Angeles, California  90024} \\  
\r 5  {\eightit University of Chicago, Chicago, Illinois 60637} \\
\r 6  {\eightit Duke University, Durham, North Carolina  27708} \\
\r 7  {\eightit Fermi National Accelerator Laboratory, Batavia, Illinois 
60510} \\
\r 8  {\eightit Laboratori Nazionali di Frascati, Istituto Nazionale di Fisica
               Nucleare, I-00044 Frascati, Italy} \\
\r 9  {\eightit Harvard University, Cambridge, Massachusetts 02138} \\
\r {10} {\eightit Hiroshima University, Higashi-Hiroshima 724, Japan} \\
\r {11} {\eightit University of Illinois, Urbana, Illinois 61801} \\
\r {12} {\eightit Institute of Particle Physics, McGill University, Montreal 
H3A 2T8, and University of Toronto,\\ Toronto M5S 1A7, Canada} \\
\r {13} {\eightit The Johns Hopkins University, Baltimore, Maryland 21218} \\
\r {14} {\eightit National Laboratory for High Energy Physics (KEK), Tsukuba, 
Ibaraki 305, Japan} \\
\r {15} {\eightit Lawrence Berkeley Laboratory, Berkeley, California 94720} \\
\r {16} {\eightit Massachusetts Institute of Technology, Cambridge,
Massachusetts  02139} \\   
\r {17} {\eightit University of Michigan, Ann Arbor, Michigan 48109} \\
\r {18} {\eightit Michigan State University, East Lansing, Michigan  48824} \\
\r {19} {\eightit University of New Mexico, Albuquerque, New Mexico 87131} \\
\r {20} {\eightit Osaka City University, Osaka 588, Japan} \\
\r {21} {\eightit Universita di Padova, Istituto Nazionale di Fisica 
          Nucleare, Sezione di Padova, I-35131 Padova, Italy} \\
\r {22} {\eightit University of Pennsylvania, Philadelphia, 
        Pennsylvania 19104} \\   
\r {23} {\eightit Istituto Nazionale di Fisica Nucleare, University and Scuola
               Normale Superiore of Pisa, I-56100 Pisa, Italy} \\
\r {24} {\eightit University of Pittsburgh, Pittsburgh, Pennsylvania 15260} \\
\r {25} {\eightit Purdue University, West Lafayette, Indiana 47907} \\
\r {26} {\eightit University of Rochester, Rochester, New York 14627} \\
\r {27} {\eightit Rockefeller University, New York, New York 10021} \\
\r {28} {\eightit Rutgers University, Piscataway, New Jersey 08854} \\
\r {29} {\eightit Academia Sinica, Taipei, Taiwan 11529, Republic of China} \\
\r {30} {\eightit Texas A\&M University, College Station, Texas 77843} \\
\r {31} {\eightit Texas Tech University, Lubbock, Texas 79409} \\
\r {32} {\eightit University of Tsukuba, Tsukuba, Ibaraki 305, Japan} \\
\r {33} {\eightit Tufts University, Medford, Massachusetts 02155} \\
\r {34} {\eightit University of Wisconsin, Madison, Wisconsin 53706} \\
\r {35} {\eightit Yale University, New Haven, Connecticut 06511} \\
\end{center}
\date{\today}
}
\maketitle
\begin{abstract}
We have searched for
chargino-neutralino production (\chichi)
in 1.8 TeV \ppb\sp collisions,
followed by their leptonic decays
\chione\dk\chizero\lnu\psp and \chitwo\dk\chizero\lplm.
These trilepton events are expected
within a framework of the Minimal Supersymmetric
Standard Model (MSSM).
In a 19.1 pb$^{-1}$ data sample collected with
the Collider Detector at Fermilab,
no trilepton events were observed. 
Upper limits on $\sigma(\ppb\dk\chichi)\cdot BR(\chichi\dk 3 \ell + X)$
were obtained for various MSSM parameter space regions,
yielding new 95\% confidence level lower limits
for the neutralino (\chitwo) mass which extended as high as 49 \mgev.
\end{abstract}
\vspace{.2in}
\vspace{.2in}

\narrowtext

Although the Standard Model (SM) provides remarkable agreement with current high
energy physics data, it fails to provide insight into several important issues.
Among these are the apparently arbitrary energy scale of 
electroweak symmetry breaking, the appearance of divergences in the Higgs
boson self-energy~\cite{divergences}, and the failure of coupling constants to
unify at large energy scales~\cite{cconst_extrap}.  
A simple extension to the SM to solve these difficulties
is the Minimal Supersymmetric Standard Model (MSSM)~\cite{basicMSSM}.

In the MSSM, there are two charged and four neutral supersymmetric (SUSY)
partners ($\tilde{\chi}$'s) of electroweak gauge bosons and Higgs bosons.
In \ppb\sp collisions the lightest chargino (\chione) and the second lightest
neutralino (\chitwo) are pair-produced along
with their subsequent leptonic decays
\chione\dk\chizero\lnu\sp and \chitwo\dk\chizero\lplm,
in which \chizero\sp is the lightest neutralino 
(lightest supersymmetric particle or LSP) and is stable.
We expect an appreciable rate of 
the cross section times branching ratio
($\sigma \cdot BR$) for the resulting trilepton final state in the MSSM
with the Grand Unified Theory (GUT) hypothesis 
provided by Supergravity~\cite{basicSUGRA}
and slepton/sneutrino mass constraints~\cite{rge}.
The trilepton final state has small SM backgrounds, making it an
excellent discovery signature at hadron colliders~\cite{Baer1}. 

We present results of the search for
\chichi\dk$\ell^{\pm}\ell^{+}\ell^{-}+X$ events ($\ell$ = $e$ or $\mu$)
using 19.1 \invpb\sp of data
from \ppb\sp collisions at a center of mass energy of $\sqrt{s}$ = 1.8 TeV.
The data sample was collected at the Collider Detector at Fermilab (CDF) 
during the 1992-93 run of the Fermilab Tevatron.
The CDF detector is described in detail elsewhere~\cite{CDFdet}.
The portions of the detector relevant to this analysis will be 
described briefly here.
The location of the \ppb\sp collision event vertex ($z_{vertex}$)
is measured along the beam direction
with a time projection chamber (VTX). 
The transverse momenta (\PT) of charged particles are measured 
in the pseudorapidity region $|\eta|<$ 1.1 by the
central tracking chamber (CTC), which is situated in a 1.4 T solenoidal
magnet field.
Here $\PT = P \sin \theta$, $\eta$ = $\rm - \ln \tan(\theta/2)$, and
$\theta\psp$ is the polar angle with respect to the proton beam direction.
The electromagnetic (EM) and hadronic (HA) calorimeters are located
outside the tracking chambers,
segmented in a projective 
tower geometry, 
and covering the central (CEM, CHA; $|\eta|<1.1$) and
plug (PEM, PHA; $1.1<|\eta|<2.4$) regions.  
Muon identification is available in the central muon (CMU, CMP; 
$|\eta|<0.6$)
and muon extension (CMX, $0.6<|\eta|<1.1$) detectors.  

The trilepton candidates are selected from an initial sample of 
$6.3 \times 10^6\psp$ events
that have fired the inclusive central electron or muon 
triggers with $\PT\psp>$ 9.2 \pgev.
We require the events to contain at least
one lepton candidate passing strict lepton identification requirements
and at least two additional lepton candidates
with less stringent requirements.
A strict electron candidate must deposit at least 11 GeV transverse energy (\ET)
in the CEM, exhibit lateral and longitudinal shower profiles 
consistent with an electron,
and be well matched to a charged track~\cite{Photon_conversion} with 
\PT\psp$\geq$ \ET/2.
A strict muon candidate must produce a track segment in the
CMU and/or CMP chambers, be well matched to
a charged track with \PT\psp$\geq$ 11 \pgev, and deposit 
calorimeter energy consistent with a minimum ionizing (MI) particle.  
Loose electron selections accept CEM or PEM energy clusters,
whose shower profiles are consistent with an electron,
with \ET\psp$\geq$ 5 GeV.
The CEM electron is required to be well matched to
a charged track~\cite{Photon_conversion} with \PT\psp$\geq$ \ET/2, while
the PEM electron must be correlated with a high occupancy of hits in the VTX.
Loose muon selections identify 
track segments in the CMU, CMP or CMX with \PT\psp$\geq$ 4 \pgev.
In addition, a charged track with \PT\psp$\geq$ 10 \pgev\sp
outside the central chamber coverage~\cite{CDFdet} is considered
a central MI (CMI) muon 
if it deposits energy in the central calorimeters consistent
with an MI particle.

We further require:
(a)~each lepton to pass a lepton isolation ($ISO$) cut in which
	the total calorimeter \ET\sp in an $\eta$-$\phi$ cone of radius
	$R \equiv \sqrt{(\Delta\phi)^2+(\Delta\eta)^2}$ = 0.4 
	around the lepton, excluding the lepton \ET,
	must be less than 2 GeV;
(b)~$|z_{vertex}| \leq$ 60 cm;
(c)~the $\eta$-$\phi$ distance ($\Delta R_{\ell \ell}$)
	between any two leptons to be greater than 0.4;
(d)~the difference in azimuthal angle ($\Delta \phi_{\ell_1 \ell_2}$)
	between the two highest \PT\sp leptons in the event to be
	less than 170$^\circ$;
(e)~at least one \epem\psp or \upum\psp pair;
(f)~removal of events containing an \lplm\psp pair with
	invariant mass in the regions 2.9-3.3 \mgev\psp ($J/\psi$), 
	9-11 \mgev\psp ($\Upsilon$) and 75-105 \mgev\psp (\Zz).
After imposing these criteria, we are left with
zero SUSY trilepton candidate events (see Table~\ref{tab:event_yields}).

We use the ISAJET Monte Carlo program \cite{ISAJET_overview} and
a CDF detector simulation program 
to determine the total trilepton acceptance ($\epsilon^{tot}$),
which consists of geometric and kinematic acceptance, trigger efficiency,
isolation efficiency, and lepton identification (ID) efficiency.

The trigger efficiency curves for single $e$'s and $\mu$'s
are obtained from data samples which are not biased by the inclusive
lepton triggers.
These curves reach a plateau above 11 \pgev\sp
at (84.3$\pm$1.5)\% for $e$'s and (88.6$\pm$0.7)\% for $\mu$'s.
The isolation efficiencies for $e$ and $\mu$ are determined from
the second leptons in \Zz\dk\lplm\sp events
(whose underlying event activity should be similar to that in SUSY events)
where no isolation cut is imposed on the second lepton.
The isolation efficiencies are (95$\pm$1)\% for central leptons and
(80$\pm$3)\% for plug electrons. Lepton ID efficiencies are also determined
from the second leptons in
\Zz\dk\lplm\sp and $J/\psi\dk\lplm\sp$ events where
no ID criteria are imposed on the second lepton.
The values obtained from \Zz\sp and $J/\psi\sp$ events agree well, 
indicating that the ID efficiencies are independent of the lepton \PT.
The resulting lepton ID efficiencies are listed in Table~\ref{tab:lepton_id}.  

The SM backgrounds can be divided into two
classes: 
$(i)$~direct trilepton events 
	(\Wpm\Zz, \Zz\Zz, \ttb, $b\bar{b}$ and $c\bar{c}$ production)
and 
$(ii)$~dilepton (Drell-Yan, \Zz, and \Wp\Wm) plus fake lepton events.
The additional fake lepton is
an object identified as a lepton,
which does not come from the main physics process.
Each of these backgrounds is estimated using 
ISAJET and the CDF detector simulation program.

In the first category of backgrounds,
the production cross sections for \Wpm\Zz, \Zz\Zz\sp and \ttb\sp 
are taken to be
2.5 pb~\cite{VB_prod}, 1.0 pb~\cite{VB_prod}
and 7 pb (top quark mass of 170 \mgev)~\cite{top_prod}, respectively.
It should be noted that the $ISO$ distributions for $b$ and $c$ decay leptons
in ISAJET agree well with those from the CLEOQQ program
(optimized for heavy flavor decays)~\cite{CLEOQQ}.
The total expected background from these processes is
1.15$\pm$0.65 events, arising entirely from 
$b\bar{b}$ and $c\bar{c}$ production, with negligible
contributions from \Wpm\Zz, \Zz\Zz\sp or \ttb.

Since the primary mechanism of Drell-Yan, \Zz\sp and \Wp\Wm\sp productions
is the Drell-Yan process,
an accurate fake rate 
($e.g$., misidentified pions, photon conversions,
decays in flight,
$b/c$ semileptonic decay leptons from initial state radiation, $etc$.)
can be estimated
by analysing well-identified \Wpm\dk$\ell^{\pm}\nu\sp$ events
(without any restriction on jets):
(0.273$\pm$0.036)\% fake leptons per event.
The fake rate is then applied to the estimated rates of
Drell-Yan, \Zz\sp and \Wp\Wm\sp productions.
We use the Drell-Yan and \Zz\sp production cross sections
measured by CDF~\cite{CDF_Xsec_WZ,CDFPRL_Xsec_DY},
while the \Wp\Wm\sp production cross section is  taken as
9.5 pb~\cite{VB_prod}.  
We estimate these background yields 
to be 0.58$\pm$0.13 Drell-Yan events, 0.14$\pm$0.03 \Zz\sp events and
negligible contribution from the \Wp\Wm\sp process.

The total of all expected backgrounds is thus 1.9$\pm$0.7 events.
This is consistent with our observation of zero events.
                                 
There are four primary sources of systematic uncertainty
in the $\sigma\cdot BR$ measurement:
trigger efficiency; trilepton-finding efficiency; structure functions; and 
total integrated luminosity.  
The single muon trigger efficiency has the largest uncertainty ($\pm 2.7$\%),
which we conservatively use for all events.
The combined systematic uncertainty of all trilepton-finding efficiencies
(kinematic, geometric, reconstruction, 
identification, and isolation) is $\pm 12.9$\%,
mainly from the geometric and kinematic uncertainties
in the detector simulation program.
The trilepton acceptance was studied 
with the CTEQ 2L structure function \cite{CTEQ_ref} as the nominal choice 
and various other structure functions~\cite{struc_fcns}.
We take the maximum deviations from the CTEQ 2L predictions as
our systematic uncertainty: $^{+8.2}_{-1.8}$\%.  
The systematic uncertainty of the total integrated luminosity
is $\pm 3.6$\%.  
Combining these four uncertainties gives a total systematic uncertainty
in $\sigma\cdot BR$ of $^{+15.6}_{-14.4}$\%.

Based on an observation of zero trilepton events,
we set a 95\% confidence level (C.L.) upper limit of 3.1 events on
the mean number of events expected.
This result is obtained by convolving the total systematic uncertainty
of $\pm$15.6\%  (as a Gaussian smearing) with a Poisson distribution.  
Given the ISAJET prediction on $\sigma\cdot BR$\sp
we exclude a particular MSSM parameter space if:
\begin{eqnarray}
        \sigma\cdot BR(\tilde{\chi}^{\pm}_{1}
        \tilde{\chi}^{0}_{2} \rightarrow 3\ell+X)  & > &
	\frac{3.1}{ \epsilon^{tot} \cdot \int{\cal L}dt } \;\! .
\label{Eq:limit}
\end{eqnarray}
The value of $\epsilon^{tot}$ ranges from $\sim$1\% to 7\%
in the parameter region described below,
and is approximately linearly dependent on the \chione\sp mass
(40-70 \mgev).

Assuming relations of the slepton and sneutrino masses to
the gluino and squark masses~\cite{rge},
the MSSM predictions from ISAJET depend on 
the ratio of Higgs vacuum expectation values tan$\beta$,
the Higgs mixing parameter $\mu$,
the gluino mass $M(\gluino)$,
the squark-to-gluino mass ratio $M(\squark)/M(\gluino)$,
the pseudoscalar Higgs mass $M(H_A)$ and
the trilinear top-squark ($\stop$) coupling $A_t$.
The last two parameters are fixed ($M(H_A)$ = 500 \mgev, $A_t$ = 0), 
since they do not significantly alter the trilepton yield.  
Generally, allowed values of tan$\beta\psp$ are in the range 
1 to $\sim$60. 
Values close to 1 are theoretically disallowed 
(the lightest $\stop_1$ becomes the LSP). 
For tan$\beta \psp \gtsim$ 10, the 
bottom-squark ($\tilde{b}_{1}$) and tau-slepton ($\tilde{\tau}_{1}$)
can become light, due to mixing in these sectors.
Consequently, the branching ratios for 
$\chione\dk\tilde\tau_1 \nu_\tau\psp$ and
$\chitwo\dk\tilde\tau_1 \tau\psp$ increase.
Thus, the sensitivity of the search is somewhat degraded for
$\tan\beta$ values above 10.
Our trilepton sensitivity is lost for $|\mu|<$ 100 GeV 
(where the leptonic branching ratios of the chargino and neutralino 
decrease significantly),
and $|\mu|$\psp is favored to be \ltsim 1000 GeV
(the approximate energy scale below which SUSY phenomena should be observable).
Finally, the $M(\squark)/M(\gluino)$ ratio is theoretically favored
to be greater than unity~\cite{rge} and the trilepton yield
drops rapidly when this ratio exceeds 2 (this is due to sleptons becoming
heavy, which reduces the neutralino leptonic branching ratio).
Thus, we have scanned the following ranges of MSSM parameters:
$\tan\beta$ = 2, 4, 10; 
200~GeV $< |\mu| <$ 1000~GeV; 
$M(\gluino)$ = 120$\sim$250~\mgev;
$M(\squark)/M(\gluino)$ = 1.0, 1.2, 2.0.

\begin{figure}[phtb]
\epsfxsize=3.375in
\center{\leavevmode\epsffile{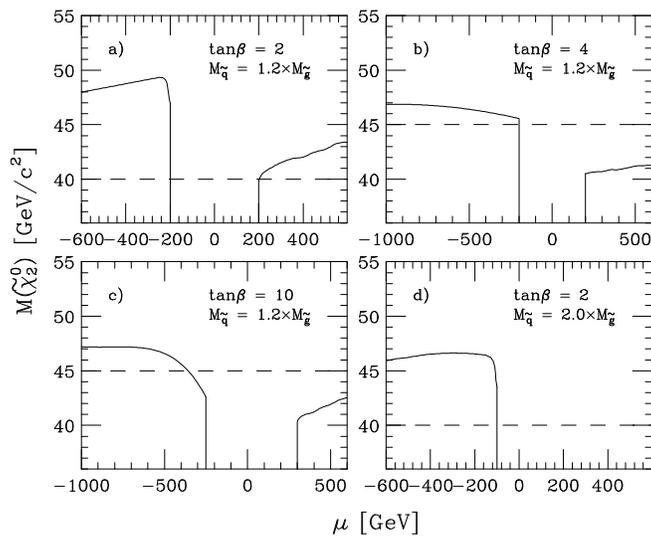}}
\caption{Neutralino (\chitwo) mass lower limits
obtained in the trilepton analysis (solid line).
The SUSY parameters used for each plot were: 
a) tan$\beta$ = 2,  $M(\squark) = 1.2 \times M(\gluino)$;
b) tan$\beta$ = 4,  $M(\squark) = 1.2 \times M(\gluino)$;
c) tan$\beta$ = 10, $M(\squark) = 1.2 \times M(\gluino)$;
d) tan$\beta$ = 2,  $M(\squark) = 2.0 \times M(\gluino)$.
The dashed line is the limit extracted
from LEP measurements \protect\cite{LEP_gaugino_limits}.
Note that $\mu\psp$ only extends down to $-$600 GeV 
for tan$\beta$ = 2.}
\label{Fig:exclusions}
\end{figure}

This analysis is insensitive to \chione\sp masses above
the current value (47 \mgev~\cite{LEP_gaugino_limits})
for any choice of MSSM parameters.
However, 
Figure~\ref{Fig:exclusions} shows several parameter space regions for 
which this analysis increases the existing \chitwo\sp mass
limit~\cite{LEP_gaugino_limits}, reaching as high as 49 \mgev\sp
at tan$\beta$ = 2.
With Equation~\ref{Eq:limit}, we also provide
the 95\% C.L. upper limits on $\sigma \cdot BR$ (single trilepton mode).
At a particular choice of the MSSM parameters
($\tan\beta$ = 2, $M(\squark)/M(\gluino)$ = 1.2, $\mu = -400$~GeV),
it is determined to be
1.4 pb, 0.6 pb and 0.4 pb for \chione\sp masses of 45, 70 and 100 \mgev,
respectively.

In conclusion, we find no events consistent with \chichi\sp pair production
in 1.8 TeV \ppb\sp collisions 
and set lower limits on the \chione\sp and \chitwo\sp masses.
The resulting \chione\sp mass limits are less than or equal to existing bounds. 
However, the \chitwo\sp mass lower limits obtained are as high as 49 \mgev\sp
in particular regions of the MSSM parameter space,
improving previous bounds~\cite{LEP_gaugino_limits}.

We thank the Fermilab staff and the technical staffs of the
participating institutions for their vital contributions.  
We also thank R.~Arnowitt, H.~Baer, and J.L.~Lopez for important discussions.
This work was
supported by the U.S. Department of Energy and National Science Foundation;
the Italian Istituto Nazionale di Fisica Nucleare; the Ministry of Education,
Science and Culture of Japan; the Natural Sciences and Engineering Research
Council of Canada; the National Science Council of the Republic of China;
the A. P. Sloan Foundation; and the Alexander von Humboldt-Stiftung.
%
%

\begin{table}
\caption{Cumulative number of events left after each cut in the 
trilepton analysis, listed
separately for the electron and muon trigger samples. 
The original CDF data sample corresponds to $\intlum = 19.1\pm0.7\psp$ \invpb.}
\label{tab:event_yields}
\begin{center}
\begin{tabular}{lcc}                                                    
Cut & $e$ triggers & $\mu$ triggers \\
\hline
Original sample          &  $3,677,903$    & $2,707,852$        \\
Dilepton events  & $ 5,472$    & $ 6,606$  \\
Trilepton events  & $    94$    & $136$  \\
\hspace{0.3cm} $ISO$ $<$ 2~GeV  &          5 &           21 \\
\hspace{0.3cm} $|z_{vertex}|<$ 60 cm  &     5  &           21                \\
\hspace{0.3cm} $\Delta R_{\ell \ell} >0.4$  &       3   &    2 \\
\hspace{0.3cm} $\Delta \phi_{\ell_1 \ell_2} <170^\circ$  &   2   &    2      \\
\hspace{0.3cm} Require $\ee$ or $\mumu$     &       2   &    2 \\
\hspace{0.3cm} \Zz\sp removal (75-105 \mgev) &       0   &    1 \\
\hspace{0.3cm} $J/\psi$ removal (2.9-3.3 \mgev)&    0   &    1 \\
\hspace{0.3cm} $\Upsilon$ removal (9-11 \mgev) &    0   &    0               \\
\end{tabular}
\end{center}
\end{table}
\begin{table}
\caption{Lepton ID efficiencies ($\epsilon$) obtained from
	\Zz\dk\lplm\sp and $J/\psi\dk\lplm\sp$ events in CDF data.}
\label{tab:lepton_id}
\begin{center}
\begin{tabular}{l c | l c}
 Muon type & $\epsilon$ (\%) & Electron type & $\epsilon$ (\%) \\ 
\hline\hline
Strict CMU and CMP  & 89.0$\pm$2.6 & Strict CEM  & 82.5$\pm$1.5 \\
Loose CMU and CMP   & 93.5$\pm$2.0 & Loose CEM   & 85.0$\pm$1.4 \\
Loose CMX           & 94.0$\pm$2.9 & Loose PEM   & 89.0$\pm$1.5 \\
Loose CMI           & 92.5$\pm$4.2 & 		 &		\\
\end{tabular}
\end{center}
\end{table}

\begin{thebibliography}{99}
\bibitem{Photon_conversion}
	We removed a track consistent with a photon conversion
	($\gamma \rightarrow e^+ e^-$).
\end{thebibliography}
\end{document}